# Surface characterisation of template-synthesised multi-walled carbon nanotubes

M.-R. Babaa [a,*], E. McRae [a,*], S. Delpeux [b], J. Ghanbaja [a], F. Valsaque [a], F. Béguin [b]

[a] *Laboratoire de Chimie du Solide Minéral, UMR-CNRS 7555, Université Henri Poincaré-Nancy 1, BP 239, 54506 Vandœuvre-lès-Nancy, France*

[b] *Centre de Recherche sur la Matière Divisée, CNRS-Université, 1B rue de la Férollerie, 45071 Orléans cedex 02, France*

Abstract

Physisorption studies and transmission electron microscopy have been used to characterise multi-walled carbon nanotubes (MWNTs) made by a template-synthesis technique. Microscopic investigations revealed formation of 'branched nanotubes' with sig- nificant irregularities in diameters and with structural defects on the external surfaces of the tubes. Krypton adsorption isotherms at 77 K were recorded; comparison of these isotherms with those obtained under the same conditions on well defined MWNTs made by the catalytic chemical vapour deposition (CCVD) technique is discussed in the light of the sample morphologies. The effect of annealing on the crystallinity of the surface is reported.

1. Introduction

Since their discovery in 1991, carbon nanotubes have attracted much interest for their potential applications in various fields such as nano-electronics [1], gas and energy storage [2,3] and molecular sieves [4]. Such quasi- one-dimensional systems are also of much fundamental interest.
Multi-walled carbon nanotubes (MWNTs) can be obtained using various methods [5], such as electric arc dis- charge, laser ablation or catalytic decomposition of hydrocarbons. While each technique has its own advantages, their common disadvantages are essentially the large distribution in tube diameters and the subsistence of impurities despite purification steps. These both constitute serious impeachments to using nanotubes as technological materials.

* Corresponding authors. Fax: +33 383 68 46 15.
*E-mail addresses:* rachid.babaa@lcsm.uhp-nancy.fr (M.-R. Babaa), Edward.McRae@lcsm.uhp-nancy.fr (E. McRae).

The template-synthesis technique using an anodic aluminium oxide film (AAO) [6], presents the advantage that nanotubes can be synthesised without any catalyst since the porous anodic alumina film acts as both a catalyst and a template during the formation of the nano- tubes. Furthermore, this offers the real possibility of synthesising tubes with uniform diameter and length and controlling the wall thickness. However, many studies have shown that such a synthesis procedure leads to the presence of structural irregularities [7,8] related to the experimental conditions.

Carbon nanotubes are attractive for low-pressure physisorption studies, since their surface is closely related to that of graphite, which is considered as a reference substrate for investigations on two-dimensional ad- sorbed phases [9]. They offer the opportunity of following the dependence of the adsorbate properties on the curvature of the graphene planes and on the confinement of the molecules within the one-dimensional channels constituted by the insides of the narrower nanotubes.

Krypton adsorption isotherms measured on exfoliated graphite exhibit vertical steps, representative of successive monolayer condensations [9]. The step pressures are determined by the substrate attraction forces which decrease with increasing distance from the surface. Such stepwise isotherms, on uniform patches of the surface of CCVD-produced MWNTs, were also strongly expected [10,11]. Comparison between adsorption isotherms measured under the same conditions on nano- tubes and on graphite allowed determining the dependence of the adsorption and wetting properties
on the specific morphology of the nanotubes [10].

This work presents a study of krypton adsorption on template-synthesised MWNTs which have, in our case, a particular morphology. The results are compared with those obtained on well-defined CCVD–MWNTs and are discussed in the light of TEM observations.

2. Experimental

   a. Sample preparation

   *Template-synthesised nanotubes*

The template-synthesis method [6] is based on two steps. First, a substrate material, i.e., a high-purity aluminium sheet (95%, 0.5 *l*m thick), was anodised in a sulphuric acid solution (20 wt%) at 0 °C, at a constant applied voltage of 15–25 V, for 5–8 h. This AAO was then used as a one-dimensional template. The deposition of pyrolitic carbon on the insides of the straight channels was carried out by exposing the AAO film to propylene at 800 °C. The nanotubes were obtained by dissolving the anodic oxide film in HF. After repeated washings in distilled water followed by filtra- tion, one of the two samples was then annealed for 1 h at 2400 °C under flowing argon.

   *CCVD-synthesised multi-walled carbon nanotubes*

The nanotubes were prepared by catalytic decomposition of acetylene diluted in nitrogen at 600 °C, according to the conditions described elsewhere [10]. NaY zeolite was impregnated by cobalt nitrate in order to ob- tain a 2.5 wt% ratio of metal/support. Once the nano- tubes were formed, the catalyst was eliminated by dissolution in 40% HF. After repeated washings in dis- tilled water followed by filtration, the remaining carbon phase was annealed for 1 h at 2400 °C under flowing argon.

   b. TEM observations

TEM studies were performed on a CM 20, 200 kV Philips apparatus with an unsaturated $LaB_6$ cathode. The samples were prepared by dispersing the powder in ethanol. After about 10 min sonication, drops of the suspension were deposited on a holey carbon copper grid.

*c. Volumetric measurements*

The apparatus used for the determination of adsorption isotherms comprises two parts. One includes the sample cell containing the adsorbent which is maintained at a constant temperature, uniform to within 0.05 K during the experiment. The other part, which is isolated from the cell by a valve, is entirely at room temperature. The manometers used for the pressure determinations are connected within this part. The gas is first introduced into the latter part and then into the adsorption cell by opening the valve. The pressure thus decreases and reaches a limiting value at equilibrium. The amount adsorbed is then determined by using the perfect gas equation which leads to the first point of the isotherm. The following points are determined using the same protocol. The manometers used are a McLeod and a Datametrics pressure gauge allowing pressure measurement between $10^{-4}$ and 1300 Pa. Thermal transpiration was taken into account by using the semi- empirical equation of Takaishi and Sensui [12]. Before each experiment, the sample was outgassed to a pressure lower than $10^{-4}$ Pa for at least 6 h at 873 K.
Krypton was chosen as adsorbate because of its low saturated vapour pressure at 77 K (225 Pa) compared to that of nitrogen ($10^5$ Pa) at the same temperature. Thus, we have good precision on pressure measurements since we work with samples of typically 20 mg.

3. Results and discussion

*a. Sample morphology*

Three samples were examined: template samples be- fore and after annealing and a CCVD sample after annealing. Figs. 1 and 2 show TEM photos of the template-synthesised MWNTs used, before and after annealing. All tubes present a number of structural 'branches' (Figs. 1 and 2a) and a number of them are irregular in diameter (Fig. 2b). Such structure has been reported elsewhere [7,8]. The tubes present different wall thicknesses, which may be explained by a non-uniform flow of the hydrocarbon on the AAO-templates or by a range of diameters in the template itself. The outer diameters of the tubes vary from 10 to 30 nm. Fig. 2c shows the presence of nanofibre-like structures in which the 0 02 planes are approximately stacked along the direction of the fibre axis. The nanofibre shows good crystallinity based on the 002 fringes. Some of the tubes are grouped together (Fig. 2d); however, the amount of grouping in the initial sample is difficult to estimate since the samples were sonicated before the TEM experiments. TEM images taken before annealing reveal that a majority of the tubes are uncapped; however, a number of them become closed after the heat treatment. Fig. 3 shows the CCVD-sample. These tubes are longer, well separated from each other and have a diameter varying between 8 and 35 nm.

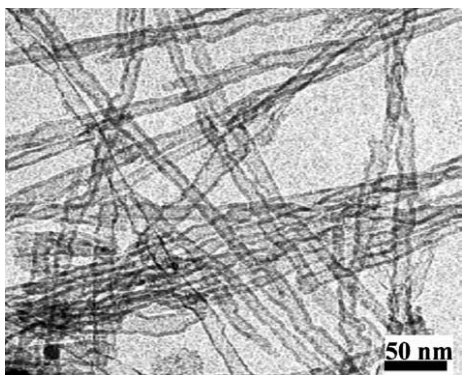

Fig. 1. TEM image of as-produced template-synthesised-MWNTs.

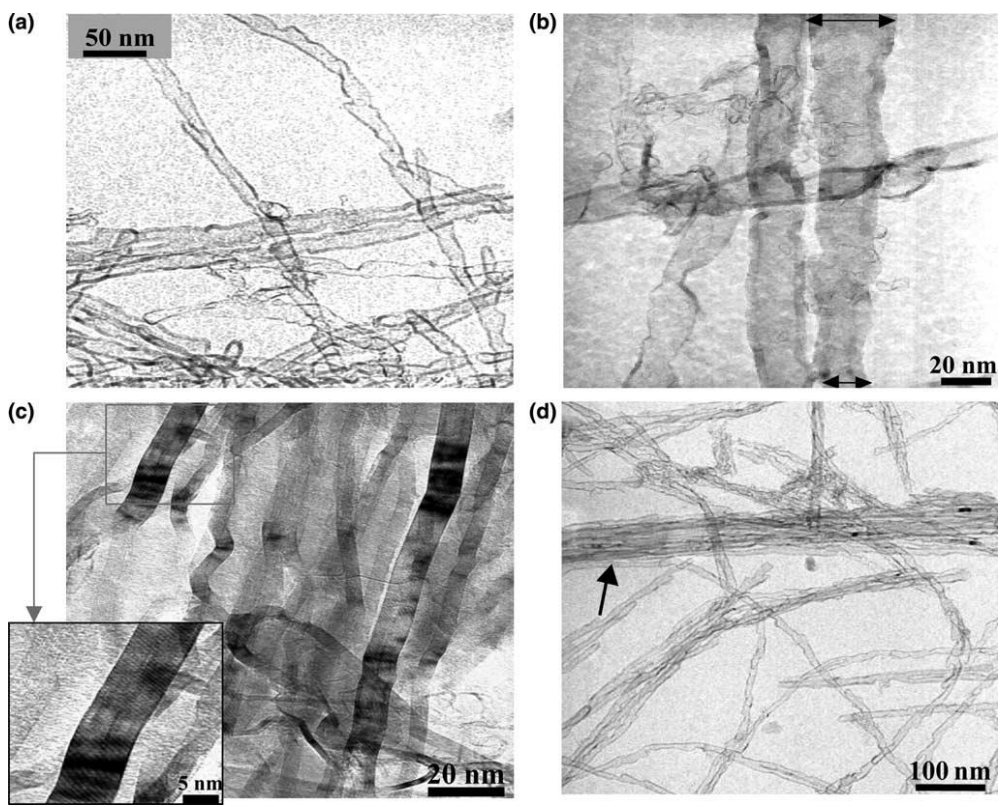

Fig. 2. TEM images of template-synthesised MWNTs after annealing.
Arrow in (d) shows a bundle-like arrangement.

    *b. Krypton adsorption on MWNTs*

The adsorption isotherm on the annealed CCVD– MWNTs measured at 77.3 K is shown in Fig. 4a. This curve exhibits two steps, of approximately the same height, which correspond to the successive formation of two monolayers on uniform or quasi-uniform patches of the external surface, before reaching the saturated va- pour pressure. No hysteresis could be observed between the adsorption and desorption processes. This is consistent with the transmission electron microscopy observations, according to which a majority of the observed tube extremities are closed.

Fig. 4b compares the adsorption isotherms of Kr at 77.3 K on template-synthesised MWNTs before and after annealing. The isotherm on annealed MWNTs exhibits the two steps observed in the case of CCVD–MWNTs, both corresponding, as mentioned above, to adsorption on the uniform or quasi-uniform patches of the external surface, namely the external walls of the nanotubes. These steps are not vertical nor is the plateau between the two steps completely horizontal. No hysteresis is observed between adsorption and desorption, which is consistent with TEM observations which show that a great number of tubes in the annealed sample are closed. This implies also that no capillary condensation takes place in the cavities present in the bundles formed by individual tubes as revealed by TEM images. The similarity in shape of isotherms measured on the two kinds of MWNTs after annealing (two steps at approximately the same pressure) indicates that if bundles are formed by individual MWNTs, their presence has little effect on the shape of the isotherms in contrast with sin- gle walled carbon nanotubes (SWNTs) in which the bundle organization offers very attractive sites namely the grooves and the interstitial channels on which adsorption occurs at a very low pressure [13,14]. The isotherm in the case of SWNTs shows evidence of condensation in these sites in addition to that on external convex surface of the bundles.

In the case of as-produced template-synthesised MWNTs, the step exhibits a significant slope (Fig. 4b, upper curve) compared to the annealed MWNTs. A second step cannot be distinguished. Such a second step is thought to be more sensitive to surface crystallinity due to the low adsorbent–adsorbate interactions. The in- crease of Kr adsorption at very low pressures might be assigned to the presence of microporosity and other attractive sites. The plateau between the two steps is inclined, and hysteresis is observed. This confirms the opening of some tubes before annealing as observed by TEM; however, annealing at 2400 °C leads to cap- ping of a majority of the tubes and provides a well de- fined surface reducing drastically the number of defects. The BET specific surface area has been measured for the three samples. By taking 0.147 nm$^2$ as the cross- section of Kr, we obtain $S_1$ = 101 m$^2$/g for the CCVD–MWNTs sample, $S_2$ = 72 m$^2$/g for the as-produced template-synthesised MWNTs and $S_3$ = 58 m$^2$/g after annealing. The small specific surface area of template- MWNTs can be the consequence of the differences in diameters between the two annealed samples (CCVD– MWNTs and template-synthesised MWNTs).

The similarities in the isotherms shapes (two steps with the same pressures) obtained on the two annealed samples in spite of the differences in morphology revealed by TEM observations, demonstrate that krypton, because of its small diameter, is not sensitive to the structural defects on the template-synthesised MWNTs surface. This reveals that the branched structure and surface irregularities must be constituted of relatively large crystallized fractions of the surface.

In order to compare qualitatively the two annealed samples, we normalize the isotherms by presenting the coverage fraction h (h = 1 at the monolayer completion) as a function of pressure. Fig. 5 reveals some differences in the shapes of the curves. Indeed, adsorption begins at a lower pressure (below 0.01 Pa) and up to 0.1 Pa the adsorbed amount is greater on template-synthesised MWNTs. The step itself thus appears better defined for the CCVD tubes. For the template-synthesised MWNTs, the lower starting adsorption pressure and the more smeared characteristics imply a smaller step. This height, proportional to the surface area of the uniform patches of the surface, thus implies that the sur- face of template-MWNTs is less homogeneous than that of CCVD–MWNTs tubes. This is assigned to the irregularities and defects in addition to some grouping together of the tubes which can provide very attractive sites. On the other hand, the second step does not occur at the same pressure. Indeed, the presence of some opened tubes in addition probably to some mesoporous texture generated by the aligned organisation, as ob- served by TEM (cf., Fig. 2d), can be at the origin of this difference between the two curves in this pressure region.

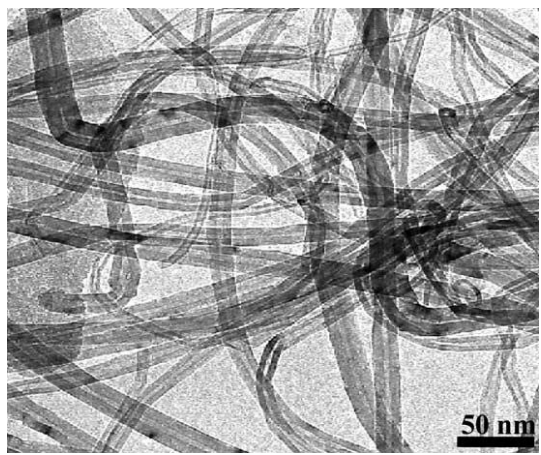

Fig. 3. TEM image of CCVD–MWNTs before annealing at 2400 °C.

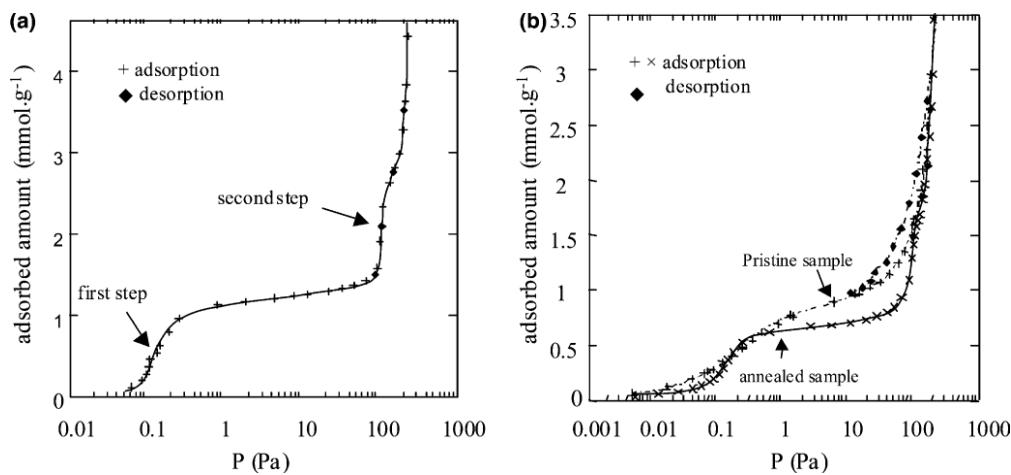

Fig. 4. Krypton adsorption at 77 K on (a) CCVD–MWNTs, and (b) template-synthesised MWNTs.

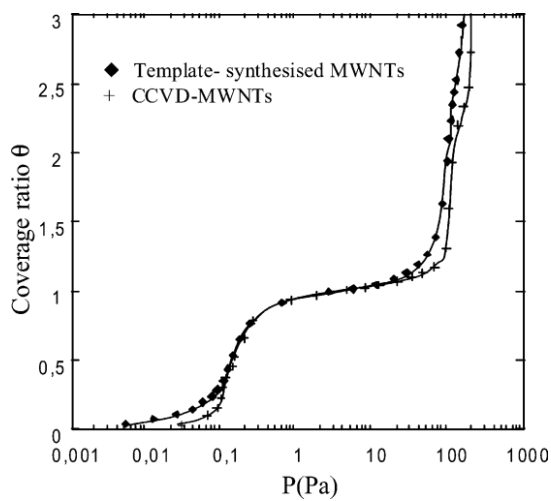

Fig. 5. Krypton adsorption at 77 K on annealed CCVD and template MWNTs.

4. Conclusions

Two samples of MWNTs of comparable average out- er diameter synthesised by different techniques have been studied by TEM and by krypton physisorption. TEM observations reveal that CCVD–MWNTs have a well defined tubular morphology; by contrast, the template-synthesised MWNTs present a number of irregularities on the surface and in the tube diameter.
Krypton adsorption on the two kinds of tubes suggests the same differences, as manifested by their specific surface areas. However, the surface irregularities do not significantly affect the condensation pressures of the two monolayers. We conclude that krypton is not sensitive to surface irregularities because of their small diameter.

Acknowledgements

We gratefully acknowledge the financial support from ''Region Lorraine'' and the ADEME (ADEME convention No. 01 74042). We are grateful to X. Duval for many fruitful discussions.

References

1. S.J. Tans, R.M. Verschuere, C. Dekker, Nature **393** (1998) 49.
2. A.C. Dillon, K.M. Jones, T.A. Bekkedahl, C.H. Kiang, D.S. Bethune, M.J. Heben, Nature **386** (1997) 377.
3. E. Frackowiak, F. Béguin, Carbon **40** (2002) 1775.
4. Q. Wang, S.R. Challa, D.S. Sholl, J. Karl Johnson, Phys. Rev. Lett. **82** (5) (1999) 956.
5. C. Journet, P. Bernier, Appl. Phys. A **67** (1998) 1.
6. T. Kyotani, L.F. Tsai, A. Tomita, Chem. Mater. **8** (1996) 2109.
7. Z. Yuan, H. Huang, L. Liu, S. Fan, Chem. Phys. Lett. **345** (2001) 39.
8. Y.C. Sui, B.Z. Cui, R. Guardián, D.R. Acosta, L. Martínez, R. Perez, Carbon **40** (2002) 1011.
9. A. Thomy, X. Duval, Surf. Sci. **299/300** (1994) 415.
10. I. Willems, Z. Kònya, J.F. Colomer, G. VanTendeloo, N. Nagaraju, A. Fonseca, J.B. Nagy, Chem. Phys. Lett. **317** (2000) 71.
11. A. Bougrine, N. Dupont-Pavlovsky, J. Ghanbaja, D. Billaud, F. Béguin, Surf. Sci. **506** (2002) 137.
12. T. Takaishi, S. Sensui, Trans. Faraday. Soc. **59** (1963) 2503.
13. M. Muris, N. Dupont-Pavlovsky, M. Bienfait, P. Zeppenfeld, Surf. Sci. **492** (2001) 67.
14. M.-R. Babaa, I. Stepanek, K. Masenelli-Varlot, N. Dupont-Pavlovsky, E. McRae, P. Bernier, Surf. Sci. **531** (2003) 86.